\documentclass{svproc}
 
\usepackage{comment}
\usepackage[utf8]{inputenc} 
\usepackage[T1]{fontenc}    
\usepackage{hyperref}       
\usepackage{url}            
\usepackage{booktabs}       
\usepackage{amsfonts}       
\usepackage{nicefrac}       
\usepackage{microtype}      
\usepackage{lipsum}
\usepackage{footmisc}
\usepackage{multicol}
\usepackage{slashbox}
\usepackage[libertine]{newtxmath}
\usepackage{color}
\usepackage{setspace}
\usepackage{xcolor}
\usepackage{pdflscape}
\usepackage{graphicx} 
\usepackage{float}
\usepackage{algorithm}
\usepackage{algorithmicx}
\usepackage{algpseudocode}	
\usepackage{indentfirst}
\usepackage{xltabular}
\usepackage{array, makecell} %
\usepackage{xcolor}
\usepackage{pgfplots}
\usepackage{tikz}

\title{ DRDrV3: Complete Lesion Detection in Fundus Images Using Mask R-CNN, Transfer Learning, and LSTM}

\author{
Farzan Shenavarmasouleh \inst{1,3} \and Farid Ghareh Mohammadi \inst{1,3}\and M. Hadi Amini \inst{2} \and Thiab Taha \inst{1} \and Khaled Rasheed \inst{1} \and Hamid R. Arabnia\inst{1}}
\authorrunning{Farzan Shenavarmasouleh et al.}

\institute{ 
  Department of Computer Science,\\ University of Georgia, Athens, GA, 30602\\
  \and Knight Foundation School of Computing and Information Sciences  \\Florida International University, Miami, FL 33199 \\
  \email{fs04199@uga.edu, farid.ghm@uga.edu, amini@cs.fiu.edu, trtaha@uga.edu, khaled@uga.edu, hra@uga.edu}
  }
  
\begin{document}

\maketitle                        
\begin{abstract}
Medical Imaging is one of the growing fields in the world of computer vision. In this study, we aim to address the Diabetic Retinopathy (DR) problem as one of the open challenges in medical imaging.
In this research, we propose a new lesion detection architecture, comprising of two sub-modules, which is an optimal solution to detect and find not only the type of lesions caused by DR, their corresponding bounding boxes, and their masks; but also the severity level of the overall case. Aside from traditional accuracy, we also use two popular evaluation criteria to evaluate the outputs of our models, which are intersection over union (IOU) and mean average precision (mAP). We hypothesize that this new solution enables specialists to detect lesions with high confidence and estimate the severity of the damage with high accuracy.

\vspace{1em}
\textbf{Keywords:} Mask R-CNN, Transfer learning, Microaneurysms, Image Segmentation, Lesions, Exudates, LSTM
 \footnotetext[3]{Authors contributed equally to this work.} 
\end{abstract}


\section{Introduction}
\subsection{\textbf{Motivation}}  
\begin{figure*}[t]
    \centering
    \includegraphics[height=1.9in]{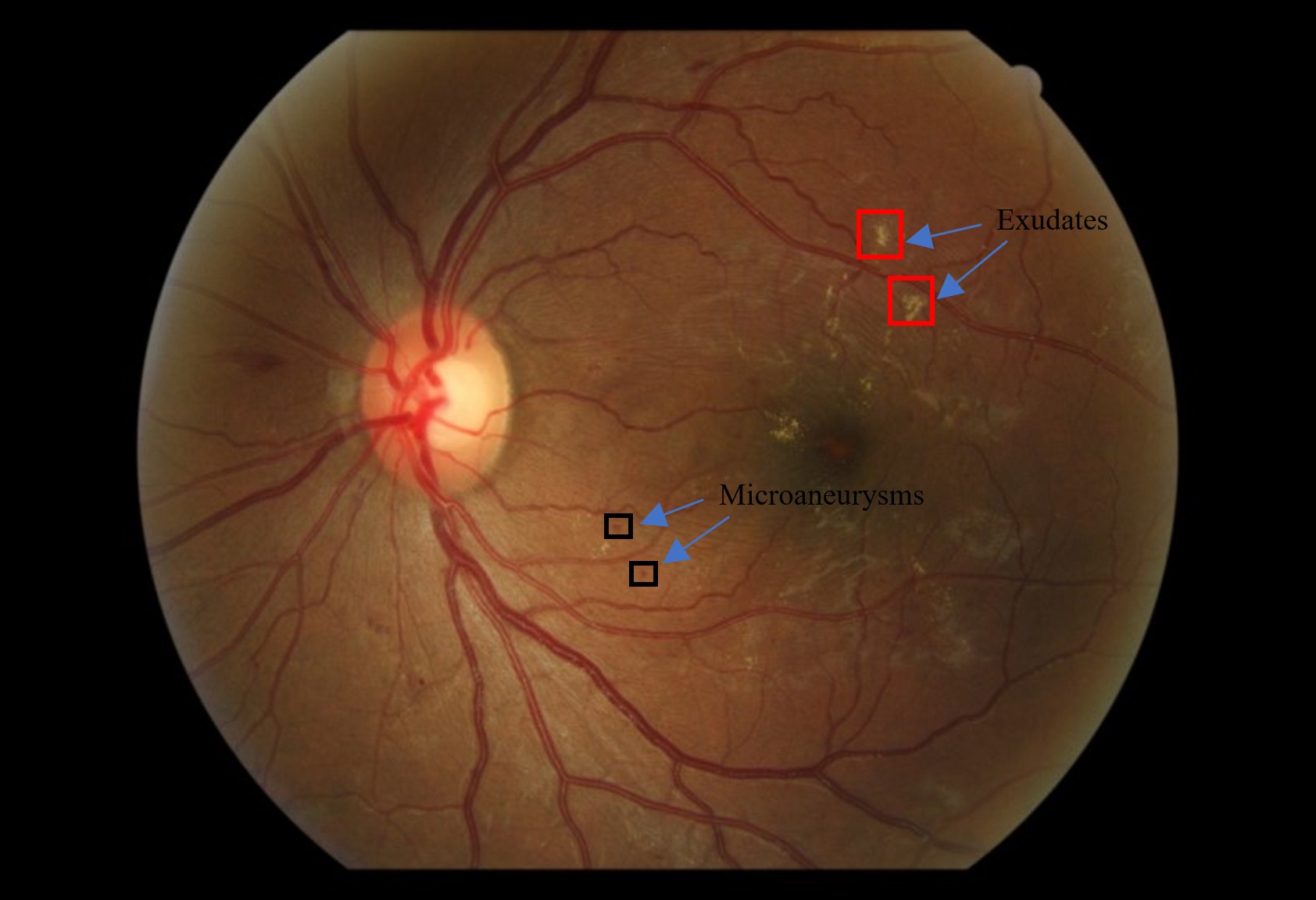}
    \caption{The lesions caused by diabetic retinopathy, namely exudates (red rectangles) and microaneurysms (black rectangles) in the eye}
    \label{exudates}
\end{figure*}

One of the infamous causes for eye damage is Diabetic Retinopathy (DR) which usually starts as vision impairment and ends up causing vision loss if not cured on time \cite{chen2019mini}. Diabetic retinopathy causes different types of vision problems in the patients’ eyes, such as cotton wool spots, exudates, microaneurysms, hemorrhages, and abnormal growth of blood vessels. In this study, we aim to focus on two of them: exudates and microaneurysms. Figure \ref{exudates} presents a sample fundus picture including exudates and microaneurysms in the eye. 

Detecting eye damages on eyes as early as possible helps specialists to have the chance to stop further injury and take advantage of laser surgeries to cure it. As the annual number of such patients is proliferated significantly and each patient needs frequent screenings, the process of analyzing the screened images is getting more important day by day. Each of the captured images requires to be technically analyzed by doctors. However, it has some limitations like time and cost. Doctors have to go through a hard and time-consuming process as the deficiencies are not always visible clearly, due to their inherent tiny sizes and unstructured shapes; hence requiring detailed assessment. Thus, the doctors have to look for an infinite number of features within each screened image. To address this inefficiency, a large number of research studies and Kaggle challenges are posted globally to make this process more automatic. Traditional accuracy, Intersection Over Union (IOU) and mean Average Precision (mAP) are examples of sample criteria that are considered in the research studies. 

Machine learning, especially deep learning plays a pivotal role in eye damage detection. Needless to say, to apply machine learning, we need to have proper datasets and leverage feature extractors, especially the ones which are compatible with the task at hand. In other words, we need to apply feature extractors which can extract important features that help machine learning algorithms to forge a model and eventually learn to predict properly. Additionally, most feature extractors are problem/task-oriented which means we cannot use an extractor trained on a certain domain for another. For instance, a feature extractor developed for detecting skin cancer \cite{sreedhar2020comparative} is not useful for detecting any other diseases.

\subsection{\textbf{Goals}} Following our work in DRDr \cite{shenavarmasouleh2020drdr} and DRDrII \cite{shenavarmasouleh2020drdr2}, our main goal here is to combine the previous networks and use the new model crafted by unifying them to 1. detect everything about lesions via a single model: lesions masks, bounding boxes, lesion types, along with the overall severity of the instances per image. 2. increase the accuracy of the severity classification that we achieved in DRDrII. 
We aspire to combine technically different image segmentation and machine learning algorithms, such as Mask R-CNN \cite{he2017mask}, LSTM \cite{hochreiter1997long}, and transfer learning, to detect lesions' attributes. We extend and evolve our Mask R-CNN model with LSTM that enables the artifact to learn more and hence tackle more tasks simultaneously.

\subsection{Challenges}
The first problem, we face in this research study is the nature of fundus pictures. They are screened and captured in different zoom levels, lighting conditions by different laboratory cameras. The images have different pixel intensity values and are collected from separate clinics with a variety of fundus photography facilities. Having these heterogeneous types of pictures, in general, negatively affect machine learning algorithms, like CNNs, and prevent the algorithm from learning and generating proper and abstract features \cite{farahani2020brief}. Hence, we need to deal with this challenge before proceeding with machine learning algorithms.

\begin{figure*}[t]
    \centering
    \includegraphics[height=1.9in]{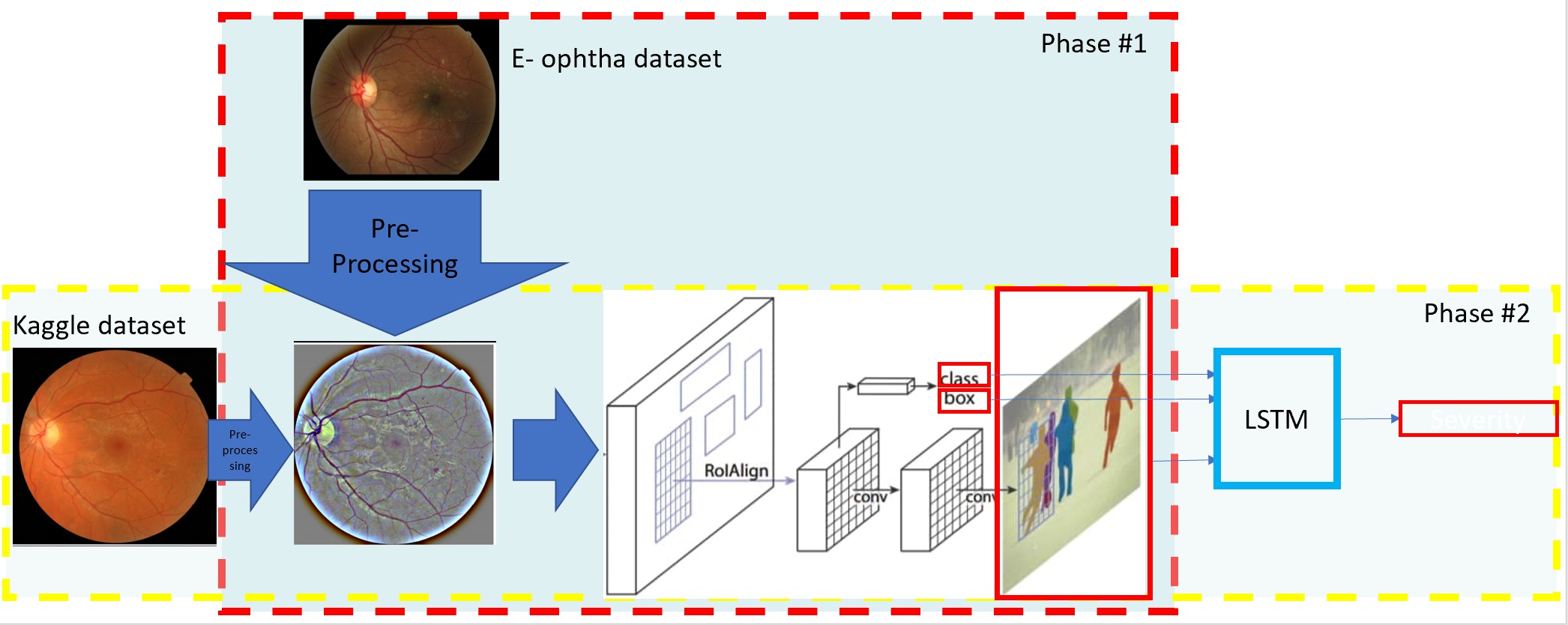}
    \caption{Proposed method in two phases}
    \label{wholephase}
\end{figure*}

The rest of this paper is organized as follows. We first discuss the related work including state of the art work including history of CNN and the latest improvement in the field. Next, we address the proposed method and problem description stating the application of transfer learning and LSTM followed by evaluation and results.

\section{Related Works}

Sopharak \emph{et al} \cite{sopharak2010machine} used a few conventional classifiers, such as Naive Bayes, Support Vector Machine (SVM), and K Nearest Neighbors (KNN) to detect one of the lesion types like exudates in the spatial level. Authors took advantage of popular methods for segmentation such as region growing  \cite{lagergren2020region} \cite{mohammadi2017region}, and thresholding \cite{pare2020image}.
The researchers in \cite{sopharak2010machine} extracted 15 features and assessed them. Additionally, the dataset they used was small with a few (39) images.

Chen \emph{et al} \cite{chen2019mini} proposed a method, so-called LFPN, to detect tiny lesion cases in diabetic retinopathy images. The LFPN architecture has two important features; first, one can use large CNN feature maps, which are an identical size to the DR image, including details of small lesion features. So, LFPN is an efficient algorithm for object classification in the second stage of RCNN. Second, one can use the top layer for region proposal, computing resources more efficiently. 

Sahoo \emph{et al.} \cite{sahoo2017automatic} established a new algorithm for Retinal Fluid Automatic Detection (RFAD) by extracting the affected retinal segment automatically and claimed to have obtained promising results. Furthermore, researchers \cite{mahiba2019severity} used hybrid color, texture features, and customized CNNs (MCNN) for a multi-class lesion classification system.

Moreover, researchers have used color to detect lesions \cite{yadav2021microaneurysm}. In \cite{yadav2021microaneurysm} researchers applied machine learning algorithms to identify microaneurysms (MAs) in fundus images using e-ophtha dataset. Among all ML classifiers, they used decision trees, Support Vector Machine (SVM) and Logistic Regression (LR), k-nearest neighbors (k-NN), Random Forest (RF), and Naive Bayes (NB) classifiers to distinguish MAs from healthy fundus images.

Researchers have used different algorithms of CNNs and schemes to tackle this problem \cite{yadav2021microaneurysm} \cite{he2017mask}. 
Several research studies are accomplished to extend the level of classification and drag the level of severity to this task too. For example, researchers in \cite{shenavarmasouleh2020drdr} presented a new method called DRDr to generate segmentation masks for two popular types of eye lesions: exudates and microaneurysms. Additionally, Shenavarmasouleh \emph{et al} in \cite{shenavarmasouleh2020drdr2} established DRDr2 to extend the lesion detection process by generating instance masks and classifying the fundus images with respect to the severity cases. DRDr2 seeks for lesions severity manually which is considered as its main drawback. Unlike most of the previous work, in this work, we present DRDrV3 in which we automatically extract useful and important features and take advantage of LSTM to learn to perform the task of classification entirely.

\section{Proposed Method and Problem Description}
Researchers \cite{he2017mask} in 2017 established a method for object instance segmentation, namely Mask R-CNN. This method detects object instances in a given input picture efficiently and creates precise segmentation masks for each of them. We find that  Mask R-CNN can be generalized to other tasks with careful fine-tuning. 
 
Mask R-CNN returns the class, bounding box, and mask information. In state-of-the-art works, researchers use Mask R-CNN for different purposes. In this study, we aim to extend and customize this architecture to make it yield the severity of the damage caused by diabetic retinopathy to the eye in addition to the previously mentioned info. Following our work in \cite{asali2020deepmsrf} we hypothesize that this can be achieved by designing a new pipeline and expand the underlying main model. The new model will also have the benefit of being able to be trained in an end-to-end manner.

In this section, we aim to propose an optimal solution to take advantage of a combination of image segmentation and machine learning algorithms. Scientists in research studies like \cite{khan2021attributes} leveraged Mask R-CNN to do image segmentation using masked dataset and get required mask information. In this study, we propose two main phases. Figure \ref{wholephase} presents two combined phases. In the first phase, we train Mask R-CNN to learn based on the masked images. Next, in the second phase, we use the result of the first phase to carefully predict the severity of damages in each case. In general, we mainly focus on Mask R-CNN to obtain these results: class/damage type (MA vs EX), bounding box, and mask. Then, we redirect all these into an LSTM model to predict the severity of the damage as well. We discuss and elaborate details of each phase clearly in the following subsections.

In this study, we use two publicly available datasets: The first dataset we use is E-ophtha  \cite{decenciere2013teleophta} offering separate masked datasets consist of exudates and aneurysms in which the images of each dataset were manually labeled and classified by ophthalmology experts. The second dataset is a public Kaggle dataset \cite{kdataset} that includes 35 thousand fundus images taken from either left or right eye, together with a single numeric value stating the severity of the instance. We use the first dataset for training masks, bounding boxes, and labels. Then we include the severity of images from the second dataset and perform the procedures explained in the second phase.

\subsection{Pre-processing}
To address the challenges discussed above, we need to apply pre-processing algorithms and convert all pictures into a homogeneous dataset. Additionally, we morph the pictures to make all the eyes in the fundus pictures look like a circle and get rid of any unwanted noises. The sizes and contrasts of eyes in the picture should never be considered a positive or negative feature and contribute on objects detection since our goal is to only look for some tiny objects in the images. Hence, we normalize these two by resizing the images to 1024x1024 pixels and also perform the same contrast normalization technique used in DRDr \cite{shenavarmasouleh2020drdr}. By doing these steps, we ensure that the feature extractors like CNN can properly begin to learn the features in our model without having to struggle with additional complexities.

Furthermore, since microaneurysms were extremely tiny and only consisted of a handful of pixels (between 1 to 5), we dilate the corresponding masks 2 times with a kernel size of 5x5 to make sure that masks are big enough to be noticeable by the artificial network.

Since all the instances of a certain defect were shown in one single mask in the first dataset (E-ophtha), to make them usable for our algorithm, we need to generate a separate binary mask for each instance shown in the mask image. To that end, we detect and detach the instances and construct an exclusive binary mask for all of them. Let's consider if one image has M instances, we generate M separate binary masks where the ones illustrate the pixels which correspond to the defect and the zero pixels show the background. Additionally, we assign class ids to the binary masks to indicate which defect each of them depicts.

\begin{figure*}[t]
    \centering
    \includegraphics[height=1.9in]{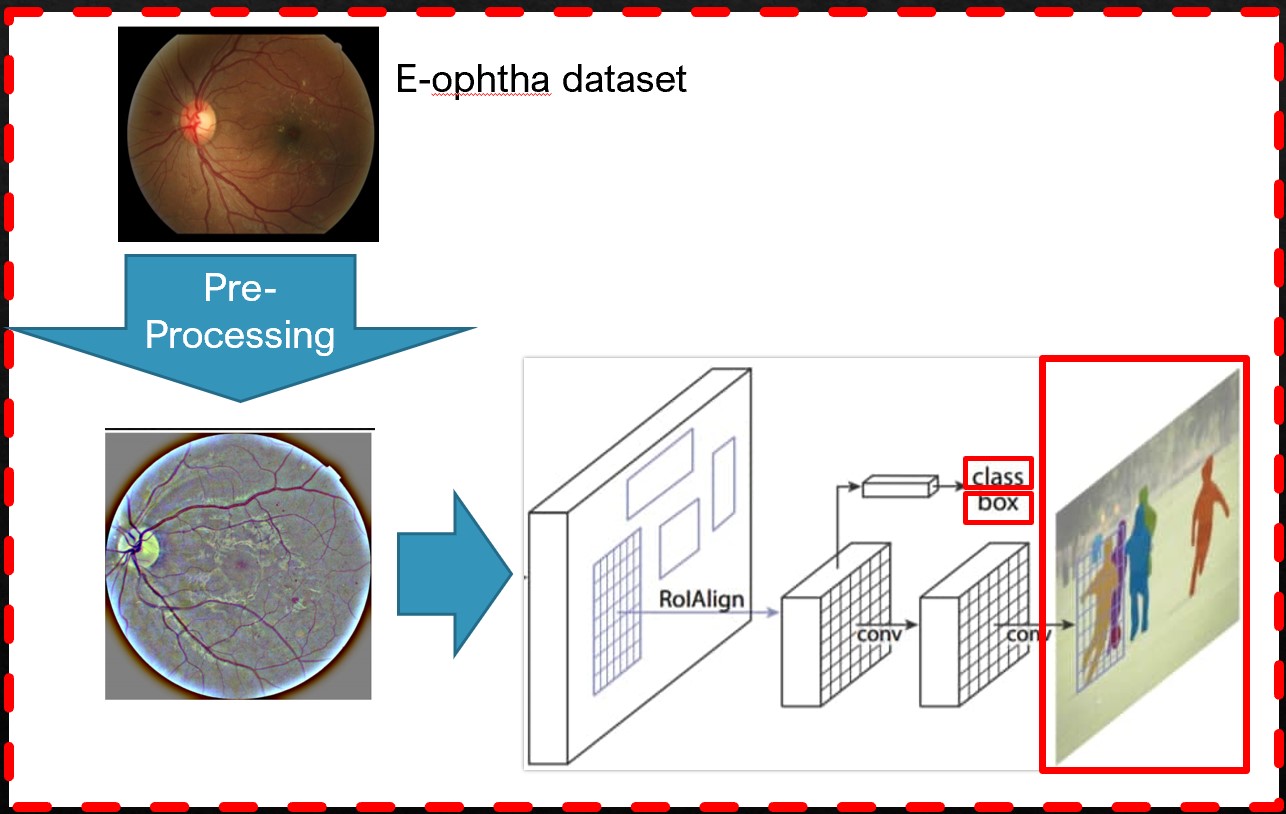}
    \caption{The architecture of phase one}
    \label{phase1}
\end{figure*}

\subsection{Phase One}
In the first phase, we aim to use E-ophtha dataset to train and evaluate Mask R-CNN performance. 

\subsubsection{Training Mask R-CNN}
In order to make Mask R-CNN work best on the dataset, we need to configure it by updating its main hyperparameters like region proposal network (RPN) sizes. For example, we decrease the size of RPN and replace the size number with as small as eight pixels. More details can be found in \cite{shenavarmasouleh2020drdr}. Furthermore, training Mask R-CNN from scratch requires a huge dataset and is time-hungry. Instead, we take advantage of transfer learning first introduced by Pan and Yang \cite{pan2009survey}, which is one of the most interesting features of deep neural networks, to address this problem with a much smaller annotated dataset.

\subsubsection{Applying transfer learning}
Due to the lack of training samples during the training phase (in phase one) which may lead out model to overfit, we employ transfer learning to address the aforementioned problem. The already existing edges in the model graph can be initialized via the weights that are publicly available via the authors of the COCO dataset \cite{lin2014microsoft}. Then we train the remaining edges (mostly edges in the last layer) while fine-tuning the previous ones if necessary. After training, our Mask R-CNN model can generate bounding boxes, masks, and assign deficit labels (EX vs MA). The performance of the model trained in this phase can be found in the Evaluation and Results section.

\subsection{Phase Two}
In this step, we take advantage of the trained Mask R-CNN in the first step, build on it to enable the model to detect the severity of damage as well, and finally evaluate this extension on a new dataset borrowed from Kaggle \cite{kdataset}.


 \subsubsection{Applying Mask R-CNN}
After executing pre-processing step, we plug in the fundus images from Kaggle dataset into the Mask R-CNN to generate the same information (bounding boxes, classes, masks for each image) we have in phase one. Furthermore, we want to add one more feature, severity, to the phase two to plug in LSTM \cite{hochreiter1997long}. 

\begin{figure*}[t]
    \centering
    \includegraphics[height=1in]{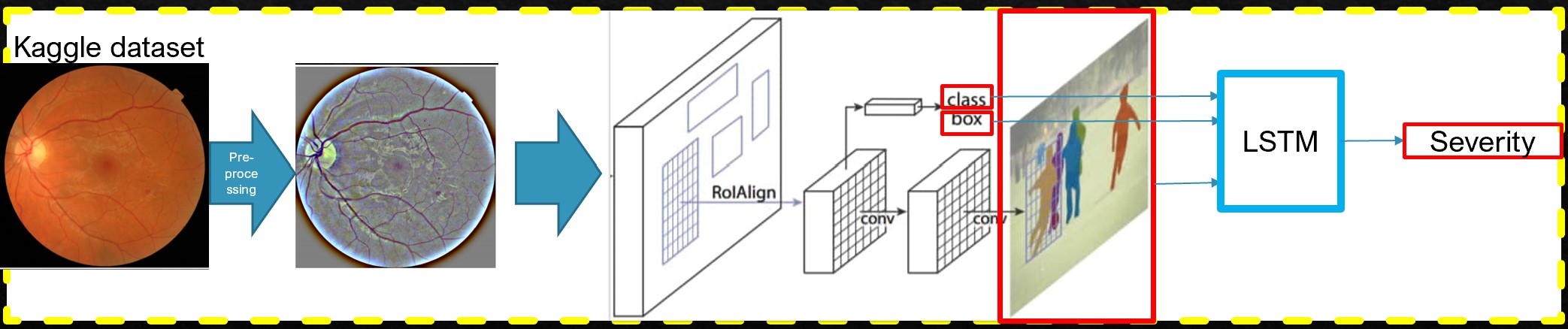}
    \caption{The architecture of phase two}
    \label{phase2}
\end{figure*}

\subsubsection{LSTM}
We aim to take advantage of LSTM \cite{hochreiter1997long} due to various reasons. LSTM does not require tuning main parameters due to the large number of its parameters such as learning rates, input, and output biases. More importantly, LSTM can handle dynamic input sizes and since we will have a different number of deficit instances generated for each image in our task, it comes to the rescue. Furthermore, LSTM helps us monitor and analyze the possible relation between sliding windows to manage the lesions associated with the locations in the image.

\begin{figure*}[t]
    \centering
    \includegraphics[height=2.0in]{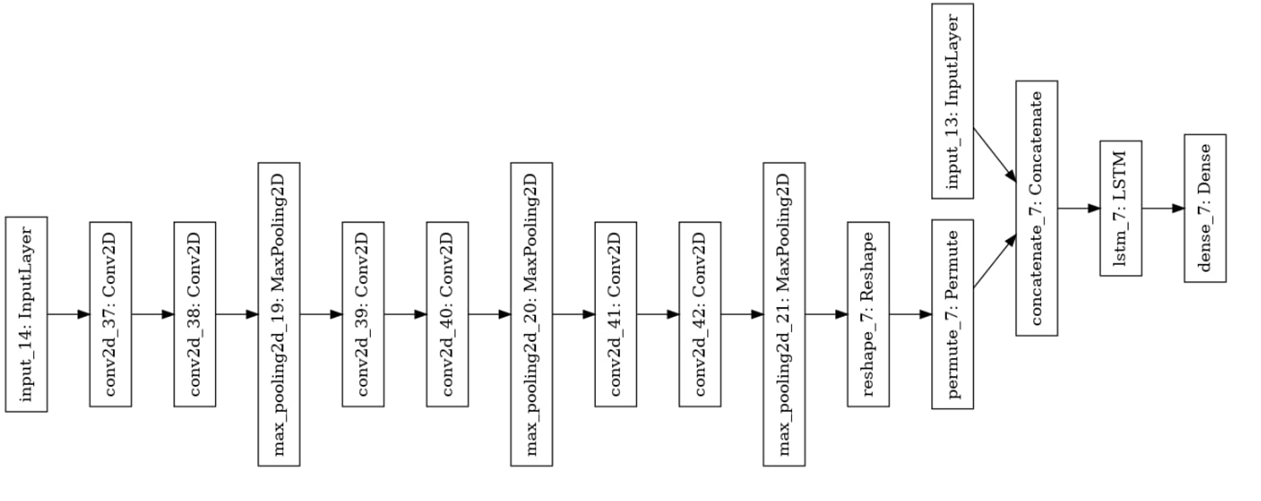}
    \caption{The input layers of LSTM}
    \label{LSTM}
\end{figure*}

In order to enable LSTM to yield the severity of each image, we need to prepare its input layer properly to combine layers of information that we have obtained from phase one. Figure \ref{LSTM} presents how input layers are connected to each other. \textit{input$\_$13: Input$\_$layer} is the concatenation of bounding boxes and class types that is being added to \textit{permute$\_$7: Permute} that is the mask layers that have been fed to 3 phases of CNN in order to extract any possible information from the shape of the masks and then reshaped to match the dimension of the first branch.


 \section{Evaluation and Results}
To evaluate our proposed method, we need to determine relevant datasets, together with evaluation criteria. In this section, we discuss both followed by the results.
 
 \subsection{Datasets}
In this study, we use two publicly available datasets: The first dataset we use is e-ophtha  \cite{decenciere2013teleophta} offering separate masked datasets consist of exudates and aneurysms where each dataset was manually labeled and classified by ophthalmology experts. E-ophtha EX contains 47 pictures with EXudates and 35 pictures are lesion-free, while e-ophtha MA provides 148 pictures with microaneurysms or small hemorrhages and 233 pictures are lesion-free.

The second dataset is a public Kaggle dataset \cite{kdataset} that includes 35 thousand fundus images taken from either left or right eye, together with a single numeric value stating the severity of the instance. It is large enough for our task to ensure that overfitting would not occur and that our result could be reproducible \cite{Shenavarmasouleh2019CausesOM}. It is important to note that this dataset does not come with any kind of mask and hence it could not be directly used in our paper without the help of the first phase.

\subsection{Evaluation criteria}
\noindent $\diamond$ \textbf{Intersection over
Union (IOU):} IOU is the fraction of the area of the overlap that happens between the actual mask and the generated (predicted) bounding box, over the area of the union of those two. When IOU=1, it means that we have a completely correct bounding box and mask. In practice, to accept the predictions a threshold is set and if the IOU positions above it then the predictions are marked as correct and incorrect the other way around \cite{jiang2020elixirnet}.

\noindent $\diamond$ \textbf{mean Average Precision (mAP):} Let's first define precision (P) which calculates the percentage of correct predictions out of all predicted bounding boxes in the image.  Average Precision (AP) presents the performance of the model for each class while detecting objects in a dataset. Finally, Mean average precision is examined as the average or mean of the AP across all of the images in a dataset \cite{bhatti2020multi}.

\noindent $\diamond$ \textbf{Accuracy:} calculates the percentage of correct predictions out of all predictions. The accuracy formula is defined in Eq. 1 \cite{sandag2020prediction}.
 
 \hspace{1in}$   Accuracy=\frac{TP+TN}{TP+FP+TN+FN}  \hspace{2in}(1)$
 
where TP is the true positive, TN is the true negative, (TP+FN) is the number of total positive predictions, FP is the False Positive, and (TN+FP) is the number of total negative predictions.
 
 \subsection{Phase One Results}
To evaluate the performance of phase one, as stated in \cite{shenavarmasouleh2020drdr}, we use mAP to compare the results with different thresholds. We use three different thresholds for IOU: 35, 50, and 75. 35 is the minimum value for IOU to mark true predictions as the model is configured previously to yield masks that have a confidence score of more than \%35. These thresholds are considered as evaluation criteria to compute mAP for each instance in the images. Table \ref{tb-phase1} presents 
 \begin{table}[th]
    \centering
    \begin{tabular}{|p{25mm}|p{25mm}|p{25mm}|p{25mm}|}
    \hline 
             test option    &mAP$\_$35 &mAP$\_$50 &mAP$\_$75  \\\hline 
       train & 0.5408 &0.5217& 0.3032\\\hline 
        Validation & 0.5113 &0.4780& 0.2563\\\hline 
         test & 0.4562 &0.4370& 0.2071\\\hline 
      
      
    \end{tabular}
    \caption{result of phase one with three different thresholds. The numbers (35,50 and 75) are the threshold of Intersection over Union (IOU)}
    \label{tb-phase1}
\end{table}

 \subsection{Phase Two Results}
In this phase, we aimed to predict the overall severity of the case in the image. We have three classes: 0 which indicates healthy, 1 that suggests medium damages, and 2 which denotes severe damages. Table \ref{tb-phase2} presents the accuracy of the test scenarios with their associated confusion matrices. The results show that the test option with classes, bounding boxes, and masks yields the best result in comparison with others and we achieve slightly better performance than our previous result in DRDrII \cite{shenavarmasouleh2020drdr2}.
 \begin{table}[th]
    \centering
    \begin{tabular}{|p{50mm}|p{25mm}|p{23mm}|}
    \hline 
             test option    &Accuracy($\%$) test & confusion matrix \\\hline 
       Classes, bounding boxes, without normalization
 & 92.67 & [2900          8       1] \hfill \break
 [72     790   20] \hfill \break
 [17    175    16]
\\\hline 
        Classes, bounding boxes, with normalization
 &84.37 &  [  2904              4             1 ] \hfill \break \hspace{0.1in}  
    [407       455       20 ] \hfill \break
 [48          145       15 ]
\\\hline 
        Classes, bounding boxes, masks
 & 93.47 &[2880   24    5]\hspace{0.1in}  \hfill \break
 [  29    771   82] \hfill \break
 [   7    111     87]
\\\hline 
      
      
    \end{tabular}
    \caption{result of phase two accuracy and the corresponding confusion matrices}
    \label{tb-phase2}
\end{table}

\section{Conclusion and Future Directions}
In this study, we bundled several machine learning and deep learning approaches and models and created a complex architecture that is capable of aiding doctors to better face Diabetic Retinopathy disease. Given a fundus image, our model can find all the instances of lesions, more specifically exudates and microaneurysms; and generate masks, bounding boxes, lesion types and finally a single severity score in an image. As our future work, we aim to explore other advanced model architectures such as U-Net. Although, U-Net is only for semantic segmentation task and our task at hand is considered instance segmentation, we want to analyze it to see if specifically, the accuracy that we achieve in phase 2 can be enhanced.

\bibliographystyle{IEEEtran}
\bibliography{bib.bib}
\end{document}